\documentclass[10pt,a4paper]{article}
\usepackage[utf8x]{inputenc}
\usepackage{ucs}
\usepackage{amsmath}
\usepackage{amsfonts}
\usepackage{amssymb}
\usepackage{amsthm}
\usepackage{algorithm}
\usepackage{algorithmic}
\usepackage{url}
\author{Wenjie Fang}
\title{New Computational Result on Harmonious Trees}
\newtheorem{conj}{Conjecture}
\newtheorem{thm}{Theorem}
\begin{document}

\maketitle

\begin{abstract}
Graham and Sloane proposed in $1980$ a conjecture stating that every tree has a harmonious labelling, a graph labelling closely related to additive base. Very limited results on this conjecture are known. In this paper, we proposed a computational approach to this conjecture by checking trees with limited size. With a hybrid algorithm, we are able to show that every tree with at most $31$ nodes is harmonious, extending the best previous result in this direction.
\end{abstract}

\section{Introduction}

In $1980$, Graham and Sloane first introduced the notion of harmonious labelling on general graphs in \cite{grahams1980additive}. For a general graph $G=(V,E)$ with $v$ nodes and $e \geq v$ edges, a harmonious labelling for $G$ is an injection $f : V \rightarrow \mathbb{Z}_{e}$ such that each edge ${x,y} \in E$ receives a distinct sum $f(x)+f(y)$ in $\mathbb{Z}_{e}$. For trees, as we have $e=v-1$, we demand $f$ to be onto instead. If $G$ allows a harmonious labelling, we say that $G$ is \emph{harmonious}.

Graham and Sloane proposed the following conjecture.

\begin{conj}
Every tree is harmonious, that is, for every tree $T=(V,E)$ with $n=Card(V)$, there is an onto mapping $f: V \rightarrow \mathbb{Z}_{n-1}$ such that $g: E \rightarrow \mathbb{Z}_{n-1}, \{x,y\} \mapsto f(x)+f(y)$ is a bijection.
\end{conj}

According to Gallian's dynamic survey on graph labelling \cite{gallian2010dynamic}, there are few results on harmonious trees. Graham and Sloane proved that caterpillars are harmonious. Aldred and McKay proved in \cite{aldred1998graceful} that every tree with no more than $26$ nodes is harmonious by computing a harmonious labelling for each such tree.

In this paper, we use a similar computational approach to prove the following result.

\begin{thm}
Every tree with at most $31$ nodes is harmonious.
\end{thm}

We extend the result in \cite{aldred1998graceful} by proposing and implementing a new algorithm for harmonious labelling. This algorithm is very similar with the one in a recent work of the author \cite{fang2010computational} on graceful labelling that also improved previous result. We will discuss details of this algorithm in the following section, which will be followed by implementation details, observations and discussions.

In the following, we will use $T=(V,E)$ to denote a general tree, with $n=Card(V)$ denoting its number of nodes. We call any onto mapping $f : V \rightarrow \mathbb{Z}_{n-1}$ a \emph{labelling} of $T$, and for any $v \in V$ we call $f(v)$ the \emph{label} of $v$. We will call the accompany $g: E \rightarrow \mathbb{Z}_{n-1}, \{x,y\} \mapsto f(x)+f(y)$ the \emph{induced labelling} of $f$, and $g(e)$ the \emph{edge label} of $e \in E$. Operations on labels are in $\mathbb{Z}_{n-1}$ unless otherwise stated.

\section{Algorithm}

To find harmonious labellings, we use a hybrid algorithm consisting of three sub-algorithms, each searches in different ways. The first two sub-algorithms, probabilistic backtracking and tabu search, resemble their counter parts in the previous work of the author \cite{fang2010computational}. The third, two-stage constraint solving, is new.

All three sub-algorithms have the property that they may fail to return a labelling, but when they succeed, the returned labelling is harmonious. This is in fact favourable, since we do not know whether every tree is harmonious, and if there is one in our range, our algorithm should not be blocked.

We use the method in \cite{wright1986constant} to generate free trees with a certain size, thus trees are represented by its level sequence. Trees are rooted in one of their center, which is midpoint of any longest path. Level of each node is its distance to the chosen root, and level sequence is a canonical sequence of levels of nodes. We can identify nodes with terms in the sequence.

\subsection{Probabilistic Backtracking}

By backtracking, we mean that we try to assign labels to nodes successively in some order, and when a node has no valid label, we \emph{backtrack} to previous nodes for alternative choices.

We choose to label nodes in the order of level sequence, which has a good property that every node is linked to some node before in the sequence. When we label nodes in this order, each assignment of label fixes a certain edge label. We label the root arbitrarily.

When we try to label a node, we may have various choices, but some of them may violate the condition for harmonious labelling. We say that a label is valid for the current node to be labelled if it maintains the property of the partial labelling on nodes other than the root being an injection, and also the partial induced labelling. It is easy to see that, if we always assign valid labels in each step, we will obtain a harmonious labelling.

We set up empirically a limit on the number of backtracking to avoid long runtime due to combinatorial explosion. This limit fails the algorithm when exceeded. To avoid being trapped in local minimal, random restart, which restarts the algorithm after a certain iterations, and stochastic perturbation, which randomly swaps pairs from time to time, are also used in actual implementation.

Here is the pseudo-code of this algorithm in Algorithm \ref{algo:backtrack}.

\algsetup{indent=2em}
\begin{algorithm}
\caption{Backtracking Search}
\begin{algorithmic}
\STATE Assign random label to the root
\WHILE{iteration limit is not exceeded}
	\IF{Current node has valid label}
		\STATE Pick one at random to assign and move to the next node
	\ELSE
		\STATE Backtrack to previous node
	\ENDIF
	\IF{all nodes are labelled}
		\RETURN the obtained labelling
	\ENDIF
\ENDWHILE
\RETURN failure
\end{algorithmic} \label{algo:backtrack}
\end{algorithm}

\subsection{Tabu Search}

Finding harmonious labelling can be modelled as a combinatorial optimization problem. For a labelling $f$, which is already onto, we define the following evaluation function.

\[ Eval(f)=n-1-Card(\{ g(e) | e \in E \}) \]

We have $Eval(f)=0$ exactly when the induced labelling $g$ is an bijection, that is to say $f$ is harmonious. We want to find a labelling $f$ that gives as low value of $Eval(f)$ as possible, which turns out to be a typical combinatorial optimization problem. This is also the approach in \cite{aldred1998graceful}.

We use tabu search to solve this problem. We start from a randomly chosen labelling $f$. The replicated label does not matter, as for any harmonious labelling $f$, $f+c$ with $c \in \mathbb{Z}_{n-1}$ is still a harmonious labelling. Then we attempt to lower the value of $Eval(f)$ by swapping labels between nodes. If a swap between $u,v$ can lower $Eval(f)$, we keep the change and put $\{ u,v \}$ into a tabu list, which forbids this swap to be done again in several turns. This is where ``tabu'' comes from.

We also stop computing when the number of iterations exceeds a certain limit, with the same reason to avoid long runtime.

Here is a pseudo-code in Algorithm \ref{algo:tabu}.

\begin{algorithm}
\caption{Tabu Search}
\begin{algorithmic}
\STATE Choose an arbitrary labelling $f$ which is already a surjection
\WHILE{iteration limit is not exceeded}
	\STATE Randomly pick several pairs of nodes not forbidden
	\IF{some pair $u,v$ such that $Eval(f)$ decreases by swapping $f(u),f(v)$}
		\STATE Swap $f(u)$ and $f(v)$
		\STATE Forbid $u,v$ from being chosen in a few next iterations with tabu list
	\ENDIF
	\IF{$f$ is a harmonious labelling}
		\RETURN $f$
	\ENDIF
\ENDWHILE
\RETURN failure
\end{algorithmic}\label{algo:tabu}
\end{algorithm}

\subsection{Two-stage Constraint Solving}

Finding harmonious labelling can also be modelled as a constraint satisfaction problem (CSP) on variables $f(v), v \in V$, with the following constraints:
\[ f : V \rightarrow \{0,1,\ldots,n-1\} \]
\[ \mathit{Alldiff}(f(v), v \in V) \]
\[ \mathit{Alldiff}(f(u) + f(v) \bmod{(n-1)}, \{ u,v \} \in E). \]
$\mathit{Alldiff}$ stands for ``all different'', which means its terms should be distinct. Clearly this model expresses exactly the condition of harmonious labelling, modulo a constant. Our previous backtracking algorithm reduces to a routine randomized backtracking algorithm to solve CSP in this model. More information about CSP solving algorithms can be found in a book by Apt \cite{apt2003principles}.

However, in this form, it is not easy to implement some usual solving techniques, such as forward checking or look ahead. We try to reduce the problem to a nicer form by the following two-stage solving.

Let $L$ be the set of leaves of $T$, and we note $P(u)$ the parent of some leaf $u$. In the first stage, we find a partial labelling $f$ over $V - L$ that does not violate conditions of harmonious labelling. This can be solved with the backtracking algorithm, since we have similar situation. With this partial labelling, we can reduce the model to the following form.
\[ f' : L \rightarrow \{0,1,\ldots,n-1\} - Im(f) \]
\[ \mathit{Alldiff}(f'(v), v \in L) \]
\[ \forall v \in L, f'(v) + f(P(v)) \bmod{(n-1)} \notin G, f'(v) \notin Im(f) \]
where $G = \{ g(e) | e \in E, e \cap L = \emptyset \}$ the set of existing induced labels.

This model decouples largely dependences between all $f'(v)$. They are now linked only by an $\mathit{Alldiff}$ constraint. In this model, we can effectively perform the forward checking optimization, which shrinks domains (possible values of a variable) by removing impossible values after each fixation of a variable. This is not effective in the first model, because variables share the same domain and the valid labels of a node depends on the value of its parent.

However, the new model is also limited by the chosen partial labelling. Even if there is a harmonious labelling, we cannot find it unless we guessed its partial labelling over $V - L$ correctly. To solve this problem, we repeat this algorithm for several times. At each run, if we solved the reduced model, we obtain a harmonious labelling by extending the partial one. If we could not find a harmonious labelling in a certain number of runs, we simply return a failure.

\subsection{Hybrid Algorithm}

The principle idea of this hybrid algorithm is that we filter trees through each sub-algorithms. Only trees failing the current sub-algorithm are sent to the next. The order is two-stage constraint solving, backtracking, then tabu search. This order is determined empirically with decreasing speed and increasing possibility of finding a harmonious labelling.

If a tree fails all three sub-algorithms, it may be a good candidate of counter-example of the conjecture of Graham and Sloane.

\section{Implementation and Result}

The previous hybrid algorithm is implemented in \texttt{C++}, including tree generation. All parameters are fixed empirically. There are various changes of algorithms for optimization, but they are all minor. Using this implementation, we are able to check computationally whether some finite families of trees are harmonious.

On an \texttt{Intel Core 2 Duo T7200}, we checked that every tree with at most $31$ nodes is harmonious. We found a harmonious labelling for every tree with our hybrid algorithm. Verification of trees with exactly $31$ nodes took roughly $2800$ hours of CPU time. To obtain the result in \cite{aldred1998graceful}, only $8$ hours of CPU time is needed.

Empirically, runtime for a single tree grows exponentially with its size, but only slightly and tolerable comparing to the growth of number of trees of the same size. With more computational power, we can extend further our checking range. There is an ongoing distributed effort on the volunteer computing project yoyo@home \cite{yoyo} targeting trees with more nodes, using an improved version of our implementation.

\section*{Acknowledgement}

We thank Charlotte Truchet from University of Nantes and Fédéric Saubion from University of Angers for their inspiring discussion about modelling harmonious labelling as a CSP problem and various CSP solving techniques.

\bibliography{harmonious}{}
\bibliographystyle{plain}
\end{document}